\documentclass[usenatbib,useAMS]{arxiv}

\usepackage{graphicx}	
\usepackage{amsmath}	
\usepackage{amssymb}	
\usepackage{multicol}        
\usepackage{bm}		
\usepackage{pdflscape}	

\usepackage{setspace}\usepackage{threeparttable}
\usepackage[dvipsnames]{xcolor}

\newcommand{\Ellen}[1] {\textbf{\textcolor{purple}{#1}}}

\usepackage[T1]{fontenc}
\usepackage{ae,aecompl}
\usepackage{bm}

\usepackage{newtxtext,newtxmath}

\title{Anisotropic cosmic-ray diffusion in isotropic Kolmogorov turbulence}

\author[Reichherzer et al.]{P.~Reichherzer$^{1,2,3}$\thanks{Contact e-mail: \href{mailto:patrick.reichherzer@rub.de}{patrick.reichherzer@ruhr-uni-bochum.de}}, J.~Becker Tjus$^{1,2}$, E.G.~Zweibel$^{4,5}$, L.~Merten$^{1,2,6}$, and M.J.~Pueschel$^{7,8}$\\\\
$^1$ Theoretical Physics IV: Plasma-Astroparticle Physics, Faculty for Physics \& Astronomy, Ruhr-Universit\"at Bochum, 44780 Bochum, Germany\\
$^2$ Ruhr Astroparticle And Plasma Physics Center (RAPP Center), Germany\\
$^3$ IRFU, CEA, Université Paris-Saclay, F-91191 Gif-sur-Yvette, France\\
$^4$ Department of Astronomy, University of Wisconsin-Madison, Madison, WI 53706, U.S.A.\\
$^5$ Department of Physics, University of Wisconsin-Madison, Madison, WI 53706, U.S.A.\\
$^6$ Institute for Astro- \& Particle Physics, University of Innsbruck, 6020 Innsbruck, Austria\\
$^7$ Dutch Institute for Fundamental Energy Research, 5612 AJ Eindhoven, The Netherlands\\
$^8$ Eindhoven University of Technology, 5600 MB Eindhoven, The Netherlands}

\date{In original form 2021 Dec 22}

\pubyear{2021}

\begin{document}
\label{firstpage}
\pagerange{\pageref{firstpage}--\pageref{lastpage}}
\maketitle

\begin{abstract}
Understanding the time scales for diffusive processes and their degree of anisotropy is essential for modelling cosmic-ray transport in turbulent magnetic fields. We show that the diffusion time scales are isotropic over a large range of energy and turbulence levels, notwithstanding the high degree of anisotropy exhibited by the components of the diffusion tensor for cases with an ordered magnetic field component.
The predictive power of the classical scattering relation as a description for the relation between the parallel and perpendicular diffusion coefficients is discussed and compared to numerical simulations. Very good agreement for a large parameter space is found, transforming classical scattering relation predictions into a computational prescription for the perpendicular component.
We discuss and compare these findings and in particular the time scales to become diffusive with the time scales that particles reside in astronomical environments, the so-called escape time scales.
The results show that, especially at high energies, the escape times obtained from diffusion coefficients may exceed the time scales required for diffusion. In these cases, the escape time cannot be determined by the diffusion coefficients. 
\end{abstract}




\begingroup
\let\clearpage\relax
\endgroup
\newpage

\section{Introduction}\label{sec:1}
Scattering by magnetic field inhomogeneities is a fundamental process in cosmic-ray transport, whether it be dominated by diffusion \citep{galprop,dragon,picard, CRPropa2017} or include a component of self-regulated streaming along the background magnetic field \citep{Kulsrud1969}.

Spatial diffusion is described via the tensor
\begin{align}
\bm{\kappa} = 
\begin{pmatrix}
 \kappa_\perp & \kappa_A & 0 \\
 -\kappa_A & \kappa_\perp & 0 \\
 0 & 0 & \kappa_\parallel
\end{pmatrix} \quad
\end{align}
within a magnetic field $\textbf{B}_\mathrm{tot}=\textbf{B}+\textbf{b}$, where $\textbf{b}$ is the turbulent component and the $z$-axis of the coordinate system is aligned with the background magnetic field $\textbf{B}$. Here, we define the perpendicular and parallel diffusion with respect to the mean magnetic field, hence $\kappa_{zz} = \kappa_\parallel$ and $\kappa_{xx} = \kappa_{yy} = \kappa_\perp$ for $\textbf{B} = B \textbf{e}_z$. Spatial diffusion enters the Parker transport equation
\begin{align}
    \frac{\partial n}{\partial t} + \textbf{u}\cdot\nabla n &= \nabla\cdot(\hat{\kappa}\nabla n) + \frac{1}{p^2}\frac{\partial}{\partial p}\left(p^2\kappa_{pp}\frac{\partial n}{\partial p}\right)  \notag \\
    &+ \frac{p}{3}(\nabla\cdot \textbf{u})\frac{\partial n}{\partial p} + S\quad . \label{eq:ParkerTransport}
\end{align}
Here, $\textbf{u}$ denotes advection speed, $p$ is momentum, $\kappa_{pp}$ is the scalar momentum diffusion coefficient, and $n$ is the  phase-space cosmic-ray density averaged over the direction of particle momentum.
In most systems, the transport is anisotropic for weak turbulence levels $b \lesssim B$, where the parallel diffusion coefficient $\kappa_\parallel$ is larger than the perpendicular diffusion coefficient $\kappa_\perp$. Due to the geometry of many pertinent astrophysical objects, however, the perpendicular component often plays a decisive role in the escape times $\tau_\mathrm{esc}$ of particles from the system. Examples include those objects where the perpendicular spatial structures have distinctly smaller scales than the structures parallel to the ordered background field. For instance, in elongated jets of active galactic nuclei (AGN), the large-scale field structure is believed to be helical. Thus, the transport perpendicular to the mean-field of the jet is associated with smaller escape times than those associated with particles leaving along the magnetic field lines of the jet. The orientation of AGN jets with respect to the observer determines the escape process relevant for observation, and objects with all jet orientations, even precessing jets, are subject of investigation. The viewing angle toward Earth then determines what signatures can be seen, see e.g.\ \citep{debruijn2020ApJ...905L..13D}. Blazars are AGN seen along the jet axis, where parallel diffusion is relevant, while the perpendicular escape process can dominate over the parallel one in inclined blazars or Fanaroff-Riley galaxies (see e.g.~\cite{bbr2005APh....23..355B,tavecchio2021, Merten2021}).

There also exist regions in the Milky Way for which the escape time of charged particles perpendicular to the Galactic plane is more efficient than the time of escape along the radial direction, despite the much smaller perpendicular diffusion, due to the small scale height of the Galactic disk \citep{Gaggero2015, Evoli2017, Reichherzer2021}. 

The often complex links between magnetic-field geometry and anisotropy of cosmic-ray diffusion must be taken into account when simulating cosmic-ray transport, as they also affect observable quantities such as the energy spectra of cosmic-ray primaries (protons, heavier nuclei) and secondaries (electrons, gamma rays, neutrinos). As the diffusion tensor changes depending on the propagation regime (set by the energy, turbulence level, turbulence spectrum, intermittency, ...), these dependencies enter the transport equation and therefore the final characteristics of the observed multimessenger signatures.
In particular, the leaky-box model of the Milky Way predicts that the cosmic-ray energy spectrum observed at Earth is steepened by diffusion: In this model, it is assumed that the particle transport is dominated by scalar diffusion and that the system is in steady state, i.e.,\ $\partial n/\partial t \approx 0$. Assuming propagation in a fixed scale height, the spectrum is given by the ratio of the source spectrum $Q(E)\propto E^{-\alpha}$ and the diffusion coefficient $\kappa_i(E)\propto E^{\gamma_i}$, i.e., $N(E)\propto E^{-\alpha-\gamma_i}$ \citep{jokipii1966, Berezinskii1990, beckertjus_review2020}, with $i$ represents the perpendicular or the parallel component, depending on the geometry of the system and the dominating component of the diffusion tensor. If we better understand the relationships between the perpendicular and parallel components of the diffusion tensor, observations of one of these components can be used to determine the other one.

This paper is organized as follows. A discussion of the theoretical background of the relation between the diagonal diffusion tensor components is presented in Section~\ref{sec:background}. Section~\ref{sec:2} covers the simulation setup for diffusion coefficient and time-scale calculations. In Section~\ref{sec:3}, the parameter space in energy and turbulence levels is examined with respect to the validity of the classical scattering relation (CSR), finding that it provides a good description of numerical results. The paper is concluded by a discussion of the results in the context of perpendicular diffusion and its consequences for the escape time scales of charged particles.

\section{Components of the diffusion tensor}\label{sec:background}

The understanding of purely parallel cosmic-ray transport has been advanced significantly over time \citep{jokipii1966, jokipii1969, Giacalone1999, Casse2001, Shalchi2009, Buffie2013, Snodin2015, Reichherzer2020, Deligny2021}, but, there still are several open questions, such as what turbulence levels allow for the application of quasi-linear theory and how the turbulence spectrum and intermittency influence particle transport \citep{Shukurov2017, Friedrich2018}. Several studies were dedicated to perpendicular diffusion coefficients and their relationship to the parallel diffusion coefficient \citep{Giacalone1999, Mace2000, Casse2001, Matthaeus2003, Candia2004, DeMarco2007, Minnie2007, Fatuzzo2010, Plotnikov2011, Harari2013, Harari2015, Snodin2015, Subedi2017, Giacinti2017, Reichherzer2020, Dundovic2020}. Knowledge of the relationships between the perpendicular and parallel components of the diffusion tensor are important in determining one component with the knowledge of the other one. When measuring diffusion coefficients of cosmic rays in e.g.\, galaxies \citep{Heesen_2021}, the orientation of the galaxy plays an important role, because geometrical arguments often allow the measurement of only one of the components of the diffusion tensor.

Analytical theories must describe the components of the diffusion coefficients over a wide range of turbulence levels and energies.
A strong turbulence level ($b/B \gg 1$), for example, results in equal perpendicular and parallel components of the diffusion coefficient as the charged particles propagate through isotropic turbulence \citep{jokipii1969, Bieber1997, Giacalone1999}. 

Test particle simulations for two-dimensional, slab, or composite (two-dimensional \& slab) turbulence can be adequately described using analytical models such as quasilinear theory (QLT) (in some domains), nonlinear guiding centre theory, and unified nonlinear theory (see, e.g.,~\cite{Shalchi2020SSRv} for a review). Despite the successes of these turbulence models, for isotropic three-dimensional turbulence, tension with these theories was found at small reduced rigidities \citep{Reichherzer2020, Dundovic2020}. So far, no encompassing theory exists capturing the relation between the parallel and perpendicular components that agrees with simulation results over the whole range of turbulence levels $b/B$ for isotropic three-dimensional turbulence. 

Instead of considering diffusion models, which, due to their underlying assumptions, only apply to certain parameter ranges or exhibit significant limitations (see, e.g.,~\citet{Mertsch2020}), one may employ a more general, nonlinear theory, such as the \textit{Bieber and Matthaeus} (BAM) theory \citep{Bieber1997}. Therein, fluctuations cause particles to deviate from the ideal helices assumed in QLT, because of continuous change of the pitch angle. Therefore, particle velocities cannot be treated as being correlated over the complete trajectory, as assumed in QLT. Instead, the general, physically motivated assumption of exponentially decaying velocity correlations applies. Consequently, the velocity correlations that are crucial for determining the running diffusion coefficients $\kappa_{ii}(t) = \langle x_i^2\rangle /(2t)$ read
\begin{eqnarray}
    \left\langle v_\perp(t)v_\perp(0) \right\rangle &=& \frac{v^2}{3}\, \cos(\Omega t)\, \mathrm{e}^{-t/\tau_\perp},\\
    \left\langle v_\parallel(t)v_\parallel(0) \right\rangle &=& \frac{v^2}{3}\, \mathrm{e}^{-t/\tau_\parallel},
\end{eqnarray}
with $\langle\cdot\rangle$ being the ensemble average, while $\Omega=v/r_\mathrm{g}$ denotes the angular frequency of the unperturbed orbit, and the effective timescales for the decorrelation of the trajectories along and perpendicular to the background field are written as $\tau_\parallel$ and $\tau_\perp$, respectively. 
In contrast to the BAM model, we make no assumptions regarding the time scales for perpendicular diffusion and, in particular, do not use the Field-line Random Walk (FLRW) coefficient as a measure of the perpendicular mean-free path. Avoiding restrictive assumptions is important, since the description of perpendicular transport as FLRW is based on large parallel mean-free paths and requires the absence of resonant scattering, neither of which criterion is fulfilled in all key astrophysical environments for the particle energies considered in the present work (the effects of FLRW \citep{Sonsrettee2016, Shalchi2021} are unlikely to be completely negligible, however, a point to which we return in the discussion of Fig. \ref{fig:overview}). Gyroresonant scattering is defined in QLT as $\cos{\Theta} = l / (2\pi r_\mathrm{g})$, where $\Theta$ denotes the pitch angle, $l$ is the wavelength of a turbulent fluctuation and $r_\mathrm{g}$ denotes the gyroradius. As $-1 \leq \cos{\Theta} \leq 1$, the maximum gyroradius for which gyroresonance occur is $r_\mathrm{g} = 2\pi/l_\mathrm{max}$. This is the upper boundary of the resonant-scattering regime (RSR). For larger particle energies and thus larger gyration radii, the resonant-scattering criterion is only sustained for a fraction of possible pitch angles. This fraction decreases with $1/r_\mathrm{g}$. Beyond the RSR, the system enters the transition regime (TR), followed by the quasi-ballistic regime (QBR) as soon as the ratio $1/r_\mathrm{g}$ becomes negligible. The energy dependence of the diffusion coefficients differ significantly in these regimes \citep{Reichherzer2020}. 
The description of perpendicular transport in the RSR also has to account for the cross-field diffusion caused by resonant scattering (see \citet{Desiati2014} for a discussion). 

The diffusion coefficients can be computed according to the Taylor-Green-Kubo (TGK) formalism \citep{Shalchi2009},
\begin{equation}\label{eq:TGK}
\begin{split}
    \kappa_\perp &= \lim\limits_{t \rightarrow \infty} \int \limits_0^t \mathrm{d}\tau\, \frac{v^2}{3}\cdot \cos(\Omega \tau)\cdot e^{-\tau/\tau_\perp} = \frac{v\,r_\mathrm{g}}{3}\frac{\Omega \tau_\perp}{1+\Omega^2\tau_\perp^2},\\
    \kappa_\parallel &= \lim\limits_{t \rightarrow \infty} \int \limits_0^t \mathrm{d}\tau\, \frac{v^2}{3}\cdot e^{-\tau/\tau_\parallel} = \frac{v^2\tau_\parallel}{3}.
\end{split}
\end{equation}
Identifying the terms $\Omega \tau_i$ as $\lambda_i/r_\mathrm{g}$ (with $i = \perp, \parallel$) allows us to express the diffusion coefficients as a function of the mean-free paths $\lambda_i$,
\begin{equation}
\begin{split}\label{eq:kappa_lambda}
    \kappa_\perp &= \frac{v\,r_\mathrm{g}}{3}\frac{\lambda_\perp}{1+\lambda_\perp^2/r_\mathrm{g}^2},\\
    \kappa_\parallel &=  \frac{v\lambda_\parallel}{3}.
\end{split}
\end{equation}
The relation between the perpendicular and the parallel components of the spatial diffusion coefficient thus reads
\begin{align}\label{eq:ratios_easy}
    \frac{\kappa_\perp}{\kappa_\parallel}= \frac{\lambda_\perp}{\lambda_\parallel}\frac{1}{1+ \lambda_\perp^2/r_\mathrm{g}^2}.
\end{align}
When $\lambda_\perp = \lambda_\parallel$, this result is similar to the hard-sphere model introduced by \citet{Gleeson1969}, which is also known as the classical scattering relation (CSR) within standard kinetic theory, and which predicts 
\begin{align}\label{eq:kappa_ratios}
    \frac{\kappa_\perp}{\kappa_\parallel}= \frac{1}{1+\lambda_\parallel^2 /r_\mathrm{g}^2}.
\end{align} 
Consequently, whenever the time scales and thus the mean-free paths in the perpendicular and parallel directions are identical, Eqs.~(\ref{eq:ratios_easy}) and~(\ref{eq:kappa_ratios}) coincide, 
and the perpendicular diffusion can be expressed as
\begin{align}\label{eq:kappa_perp}
    \kappa_\perp = \frac{\kappa_\parallel}{1+(3\kappa_\parallel)^2/(c\,r_\mathrm{g})^2}.
\end{align}
(Note that for the highly relativistic particles (Lorentz factor $\gamma > 10^8$) considered throughout the study, we approximate the velocity by the speed of light.)
Previous investigations \citep{Giacalone1999} have already shown that Eq.~(\ref{eq:kappa_perp}) will not hold everywhere. 
In this paper, we will systematically investigate the parameter space spanned by particle energy and turbulence level to deduce in which this expression and the underlying assumption $\lambda_\perp = \lambda_\parallel$ hold.


\section{Numerical simulation setup}\label{sec:2}
Test-particle propagation of highly relativistic protons is simulated with the software CRPropa\footnote{{{The specific version used for the simulations is CRPropa 3.1-f6f818d36a64.}}}, a publicly available tool for simulations of cosmic-ray transport \citep{AlvesBatista2016, CRPropa2017, 2021arXiv210701631A}. 
We performed simulations of highly relativistic protons of at least several PeV,
but many conclusions apply to the general case as well, because only the ratio of gyroradius to the correlation length $l_\mathrm{c} \approx l_\mathrm{max}/5$ is relevant for the diffusion-coefficient scaling, which extends over more than two orders of magnitude ($0.04 \lesssim r_\mathrm{g}/l_\mathrm{c} \lesssim 80$). Specifically, reducing both $B_\mathrm{tot}$ and the particle energy by the same factor results in the same diffusion tensor. 

We follow the simulation setup described in \citet{Reichherzer2021}:
For the isotropic three-dimensional turbulent component, we employ a Kolmogorov-like spectrum $G(k)\propto (k/k_\mathrm{min})^{-5/3}$ for $k_\mathrm{min} \leq k \leq k_\mathrm{max}$, with $k_\mathrm{min} = 2\pi/l_\mathrm{max}$ and $k_\mathrm{max} = 2\pi/l_\mathrm{min}$, and $G(k) = 0$ otherwise. We choose $l_\mathrm{min} = 1.7$~pc and $l_\mathrm{max} \approx 82$~pc to span a large inertial range of the turbulence and we store the turbulent field on a grid with $1024^3$ grid points and a spacing of $l_\mathrm{min}/2$. We note that this adopted inertial-range Kolmogorov scaling for the turbulence is is an idealization, as interstellar turbulence is driven on many scales, from as much as 100 pc by superbubbles to kinetic scales by cosmic rays themselves \citep{Lazarian2015}. Moreover, the Alfvénic turbulence assumed here, unlike hydrodynamic Kolmogorov turbulence, is known to be highly anisotropic, even when considering compressibility effects that can generate an isotropic component \citep{Lazarian2015}.

Our simulations of particle transport in a combined turbulent field $b$ and mean field $B$ are based on the well-established TGK formalism \citep{Shalchi2009}, see, e.g.,~\citet{Giacalone1999,Casse2001,Globus2007,DeMarco2007,Minnie2007, Snodin2015, Giacinti2017, Dundovic2020, Reichherzer2021}.
We inject 2000 particles at time $t = 0$ and position $x_i = \delta(x_{i,0})$ and integrate the particle trajectories in the magnetic field by numerically solving the Lorentz-Newton equation via the energy-conserving Boris-push method \citep{boris1972proceedings}. The step size of the particles in our integrator is $s_\mathrm{step} = \mathrm{min}(r_\mathrm{g}/5, l_\mathrm{max}/20)$ in order to sufficiently resolve the gyrations and the turbulent fluctuations.

The running diffusion coefficients are determined according to $\kappa_{ii}(t) = \left<\Delta x_{i}^{2}\right>/(2\,t\,)$ by averaging over the particles. After $\kappa_{ii}(t)$ converges to a constant value, we calculate the diffusion coefficient $\kappa_{ii}$ by averaging the final running diffusion coefficients over time to reduce statistical fluctuations. In order to ensure representative results for isotropic turbulence, we perform $20-50$ 
simulations for each parameter combination, which are only distinguished by randomly changing phases of the turbulent fluctuations with the same statistical and spectral properties.

We perform simulations only for particles in the RSR, \Ellen{TR,} and QBR. The latter ranges from $r_g \gtrsim l_c$ and is characterised by good agreement between numerical simulation results and theoretical predictions. For example, an energy scaling of $\kappa \propto E^2$ is obtained here, independent of the turbulence level. This is to be expected since diffusion is no longer dominated by resonant scattering. However, the situation is different for $r_g \lesssim l_c$, where resonant scattering becomes dominant according to the resonance criterion.\\\\
However, further reducing of the gyroradii results in two effects:
\begin{enumerate}
    \item The magnitude of the fluctuations decreases according to the relation $b \propto G(k)^{1/2} \propto k^{-5/6}$, given by the Kolmogorov-type spectrum for smaller fluctuation wavelengths. As the resonance criterion connects the gyroradii with the fluctuations wavelengths,
    reducing particle gyroradii results in decreasing magnitude of fluctuations that are relevant for resonant scatterings for these particles. In addition to the decrease in the magnitude of the fluctuations, the scales of the changes of the turbulent magnetic field are considerably larger for small gyration radii, so that an effective directed magnetic field is established on the relevant scales. Thus, an effective $b/B \ll 1$ exists on small scales.
    \item According to the resonant-scattering criterion and the limited resolution of the turbulence ($l_\mathrm{min} > \mathrm{spacing}/2$), the range of pitch angles capable of resonant interactions decreases. As the needed RAM to store the turbulence on the grid, the resolution is severely constrained and limits the range of fluctuations to a few orders of magnitude. The missing resonant-interaction possibilities due to the constrained fluctuation wavelength range lead to a numerically increased diffusion coefficient. 
\end{enumerate}
Whereas effect i) establishes the prerequisites for weak turbulence, missing resonant-scattering possibilities must be taken into account due to effect ii). Since we are both interested in the influence of the turbulence level and want to keep the numerical effort tractable, we omit the low energies. This also justifies neglecting the effect of cosmic ray self confinement, which is deemed ineffective at energies above $\sim$ 100 - 200 GeV \citep{Zweibel2017}. 

\section{Relation between perpendicular and parallel components}\label{sec:3}
In this section, we carry out a comprehensive parameter study to substantially expand on the results obtained in the different investigations \citep{jokipii1969,Giacalone1999}
and extract relevant insights on the validity of the application of the Eqs.~(\ref{eq:kappa_ratios}) and (\ref{eq:kappa_perp}).
For this purpose, a parameter space over 3 orders of magnitude in particle energy and 2 orders of magnitude in turbulence level is examined, thus covering different energy regimes, such as the RSR, the QBR and the TR.

The calculation of the diffusion coefficients via the TGK mechanism described in Section~\ref{sec:2}; see Eq.~(\ref{eq:TGK}) using the temporal convergence of the running diffusion coefficient facilitates the determination of the parallel mean-free paths $\lambda_i$. Therefore, instead of determining $\lambda_\parallel$ directly, we determine it by means of Eq.~(\ref{eq:kappa_lambda}) via the diffusion coefficient $\kappa_\parallel$ as
\begin{align}\label{eq:lambda}
    \lambda_\parallel = \frac{3\kappa_\parallel}{c}.
\end{align}
This approach simplifies the analytical prediction for the ratio of the diffusion coefficient components in the CSR, see Eq.~(\ref{eq:kappa_ratios}), to
\begin{align}
    \left(\frac{\kappa_\perp}{\kappa_\parallel}\right)_\mathrm{CSR} =  \frac{1}{1+\lambda_{\parallel, \mathrm{sim}}^2 /r_\mathrm{g}^2} = \frac{1}{1+(3\kappa_{\parallel, \mathrm{sim}})^2/(c\, r_\mathrm{g})^2}.
\end{align} 
The deviation between simulation results (labelled sim) and the CSR prediction (labelled CSR) for the ratio $\kappa_\perp/\kappa_\parallel$ yields
\begin{align}\label{eq:error}
    \varepsilon &= \left| \left(\frac{\kappa_\perp}{\kappa_\parallel} \right)_\mathrm{sim} - \left(\frac{\kappa_\perp}{\kappa_\parallel} \right)_\mathrm{CSR} \right| \cdot  \left(\frac{\kappa_\perp}{\kappa_\parallel} \right)_\mathrm{CSR}^{-1} \\
    &= \left|\left(\frac{\kappa_\perp}{\kappa_\parallel} \right)_\mathrm{sim} - \frac{1}{1+(3\kappa_{\parallel,\mathrm{sim}})^2/(c\,r_\mathrm{g})^2}\right| \cdot \left(1 + \frac{(3\kappa_{\parallel,\mathrm{sim}})^2}{(c r_\mathrm{g})^2}\right).
\end{align}
\begin{figure}
	\includegraphics[width=\columnwidth]{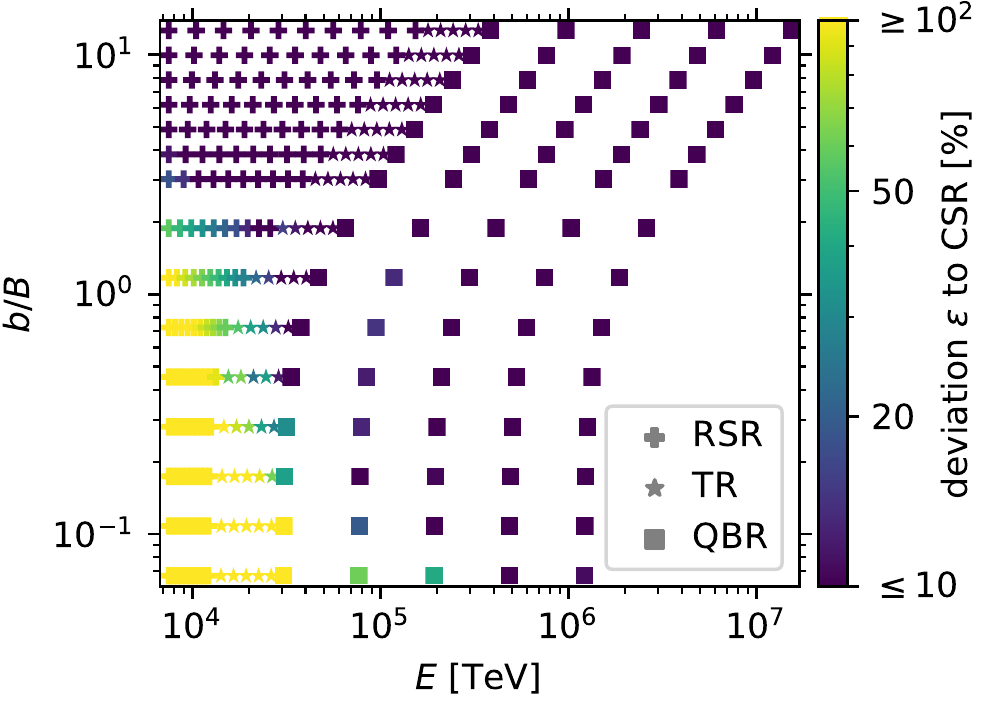}
	\caption{Comparison between simulation results and theoretical predictions from the CSR for $\kappa_\perp/\kappa_\parallel$ for all investigated simulation parameter combinations of the turbulence level $b/B$ and the energy $E$. The colour code indicates the deviation according to Eq.~(\ref{eq:error}). Each point comprises the results of 20 to 50 test particle simulations for the indicated parameter combination as described in Section \ref{sec:2}. The magnetic field strength is $B=1\,\mu$G and the maximal fluctuation wavelengths are $l_\mathrm{max}\approx 82$~pc. Using scaling laws, this plot can be applied to other astrophysical environments, as described in the text. 
	}\label{fig:overview}%
\end{figure}
Figure~\ref{fig:overview} presents this deviation $\varepsilon$ as different colours between simulated and CSR-predicted ratios of the perpendicular and parallel diffusion coefficients. The displayed data points represent all parameter combinations considered in the present study. Only for small energies and small turbulence levels is the agreement between CSR and numerical simulations poor. Particles in the zone of poor agreement satisfy $r_g < l_c < \lambda_{\parallel}$, which suggests that although these particles are scattered by  resonant waves, a significant part of the perpendicular displacement is due to FLRW, a point to which we return in the discussion of Fig. \ref{fig:sim_vs_theory_log}. A precise characterization of FLRW for our turbulence model is currently in progress, and will be described in a future publication.
Figure~\ref{fig:overview} shows that a large parameter space is well-described by this simple theoretical prediction from the CSR. Since only the ratio of the gyration radius to the correlation length is relevant for the diffusion coefficients, this plot can be used to interpret other values of $B$ by scaling the energy to the same degree as $B$. For the example of a stronger magnetic field of 1\,G, such as that in plasmoids in jets of active galactic nuclei, the same plot applies with the only change being that the energy displayed on the $x$-axis is now in units of EeV. Likewise, by reducing $l_\mathrm{max}$, $l_\mathrm{min}$ we could apply our results to lower energy cosmic rays.

As no encompassing theory currently exists capturing the relation between the parallel and perpendicular components that agrees with simulation results over the whole range of turbulence levels $b/B$ and energies for isotropic three-dimensional turbulence (see \cite{Dundovic2020}), it is not surprising that the CSR fails in describing the perpendicular diffusion for weak turbulence in the RSR. As we have shown here, an all-encompassing theory must also describe the anisotropic diffusion time scales in the RSR for weak turbulence levels with otherwise isotropic diffusion time scales even for strongly anisotropic diffusion coefficients at high energies.
Note that in many astrophysical environments (e.g., galaxies \citep{Jannson2012, Shukurov2019, Kleimann2019}), similar orders of magnitude of $b$ and $B$ are present at the injection scales of the turbulence. However, scaling of the turbulent spectra according to Kolmogorov or Kraichnan leads to $b \ll B$ on the smaller fluctuation scales relevant to lower-energy particles. In the following, we discuss the reasons for the good agreement between simulations and CSR predictions toward high energies (see Section \ref{sec:energy}) and toward high turbulence levels (see Section 2) that were identified in the overview plot.  

\subsection{Range of validity in particle energy }
\label{sec:energy}
\begin{figure}
\includegraphics[width=0.977\columnwidth]{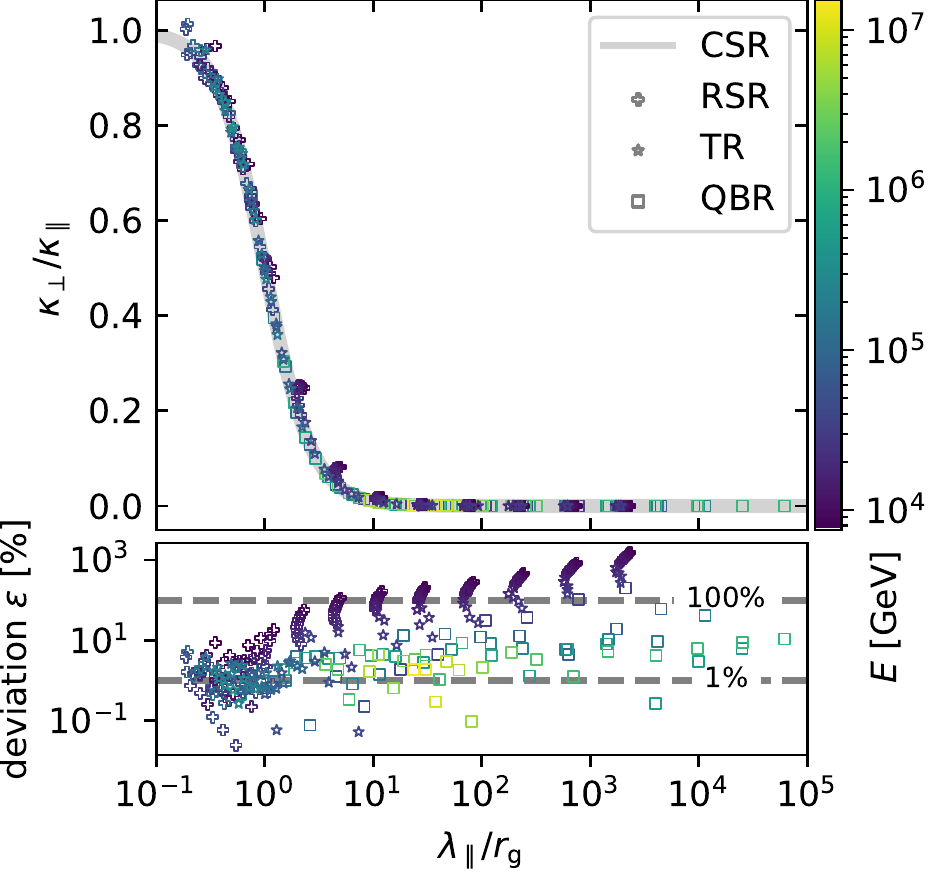}
\caption{Ratios of the perpendicular to the parallel diffusion coefficient from test particle simulations as functions of the ratio of the parallel mean-free paths and the gyroradii of the particles (top), defined in Eq.~(\ref{eq:lambda_rg}). Each point comprises the results of 20 to 50 test particle simulations and indicates via its colour-code the fixed particle energy used in these simulations. Theoretical predictions obtained from Eq.~(\ref{eq:kappa_ratios}) are shown as the grey line that is labelled theory. In the lower panel, the deviation $\varepsilon$ between simulation and theory is shown in percent and is determined via the definition presented in Eq.~(\ref{eq:error}). 
}\label{fig:sim_vs_theory_log}%
\end{figure}
Figure~\ref{fig:sim_vs_theory_log} shows simulation results for the ratios of the diffusion coefficient components $\kappa_\perp/\kappa_\parallel$ as functions of the ratio between the parallel mean-free paths and the gyroradii $\lambda_\parallel/r_\mathrm{g}$. Note that, from Eq.~(\ref{eq:lambda}), 
\begin{align}\label{eq:lambda_rg}
\frac{\lambda_\parallel}{ r_\mathrm{g}} = \frac{\kappa_\parallel}{c\,r_\mathrm{g}}.
\end{align}
The particle energy used for each simulation is colour-coded. Each data point in Fig.~\ref{fig:sim_vs_theory_log} comprises the results of 20 to 50 test particle simulations and indicates via its colour-code the fixed particle energy used in these simulations. Theoretical predictions obtained by Eq.~(\ref{eq:kappa_ratios}) are shown as the grey line labelled theory. The deviation $\varepsilon$ between theory and simulation results defined in Eq.~(\ref{eq:error}) is presented in the lower panel.

It should be noted that there is no straightforward relationship between $\lambda_\parallel/r_\mathrm{g}$ and energy, since the dependence of $\lambda_\parallel$ on energy depends on the transport regime. For this reason, the colour of the points according to the particle energy is essential to visualize the energy dependence. 
This figure, especially in the lower panel, as well as Fig.~\ref{fig:overview}, shows better agreement between theory and simulation results for higher particle energies. Particles that show large deviations ($\gtrsim 100\%$) between simulation results and CSR predictions satisfy $r_g < l_c < \lambda_{\parallel}$, as demonstrated in the lower panel of Fig.~\ref{fig:sim_vs_theory_log}. These low-energy particles (purple) correspond to small gyroradii, while the parallel mean-free paths are large due to the small turbulence level. Other definitive general statements based on these results alone are not possible because the turbulence level varies in addition to the particle energy. This will be discussed in more detail in the following subsection, where we investigate the agreement between theory and simulations for fixed turbulence levels and thus separate the influence of the turbulence level and the energy.

\subsection{Range of applicable turbulence levels}
\begin{figure}
\includegraphics[width=\columnwidth]{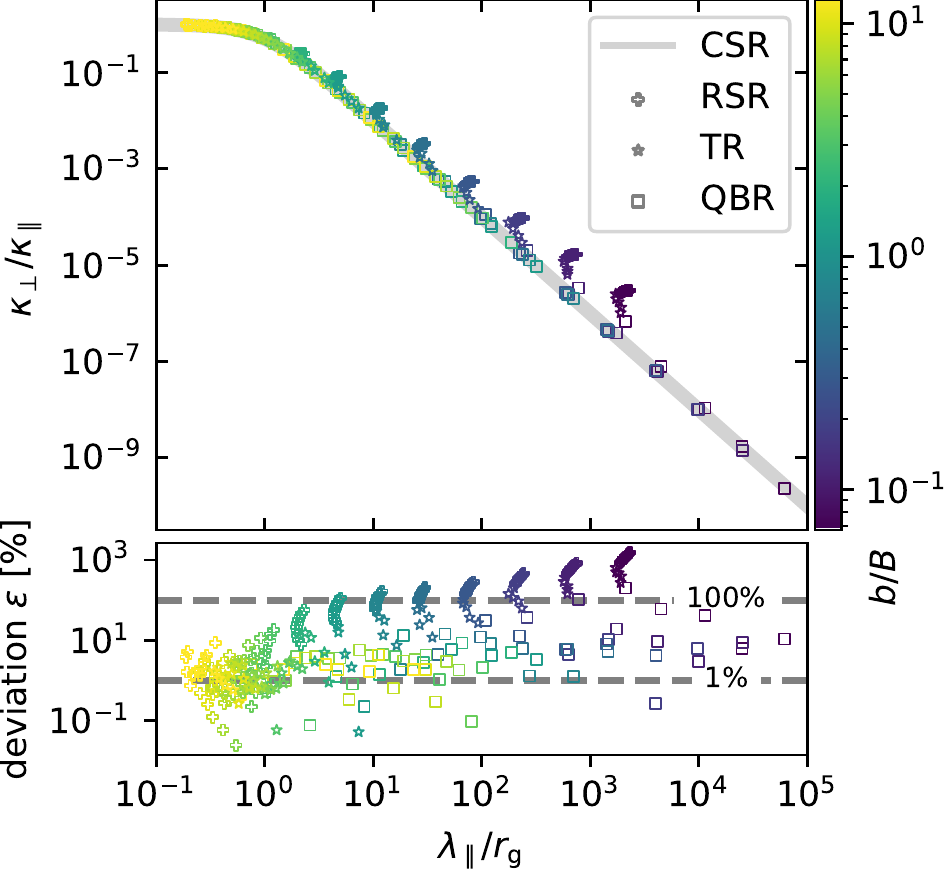}%
\caption{Same data as in Fig.~\ref{fig:sim_vs_theory_log}, except that the colour-coding indicates the turbulence level in each simulation. In the lower panel, the deviation $\varepsilon$ between simulation and theory is shown in percent and is determined via the definition presented in Eq.~(\ref{eq:error}). Good agreement between theory and simulations for high turbulence levels is found. }\label{fig:sim_vs_theory_loglog}%
\end{figure}

In order to simplify the investigation at $\lambda_\parallel/r_\mathrm{g} \gg 1$, where deviations between theory and simulations are apparent in the lower panel of Fig.~\ref{fig:sim_vs_theory_log}, the same data are shown in log-log representation in Fig.~\ref{fig:sim_vs_theory_loglog}. To enable investigation of the turbulence-level dependence of the deviation $\varepsilon$ between theory and simulation, the simulation points are colour-coded according to $b/B$.

The differences between the diagonal elements of the diffusion tensor diminish with increasing $b/B$,
so that $\kappa_\perp / \kappa_\parallel \approx 1$ in the limit $b \gg B$. Here, the ratios $\lambda_\parallel/r_\mathrm{g}$ are expected to be small, as $\lambda_\parallel$ decreases with increasing turbulence level. Since the CSR predicts exactly this, the agreement with the simulation results for this parameter range, which is characterized by strong turbulence levels, is good.

\begin{figure}
	\includegraphics[width=\columnwidth]{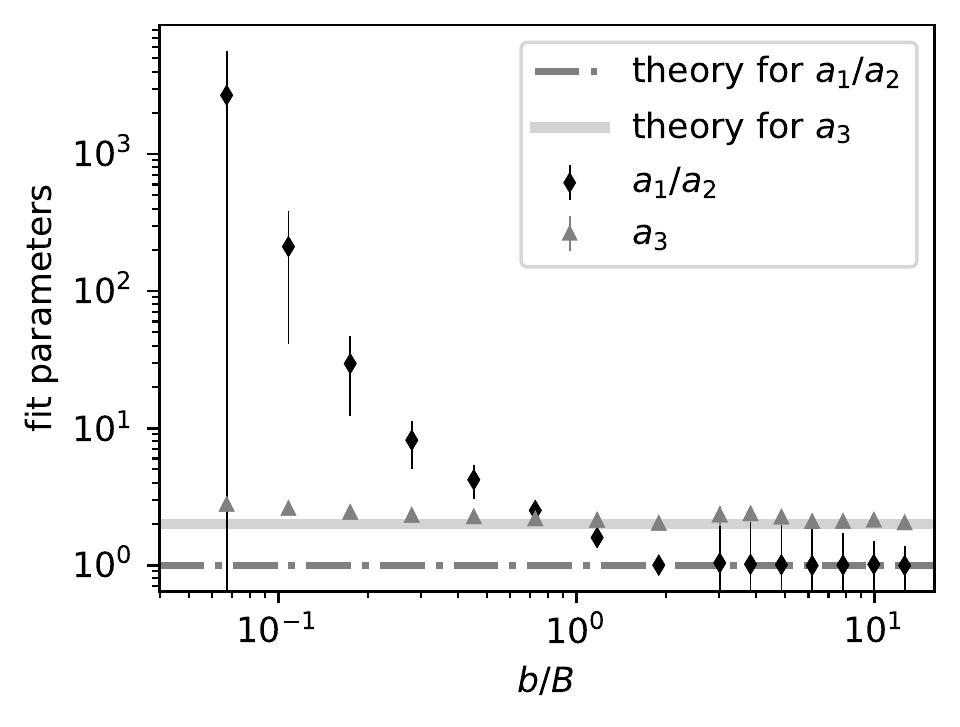}
	\caption{Fit parameters $a_3$ and the ratio $a_1/a_2$, see Eq.~(\ref{eq:fit_large}), as functions of the turbulence level $b/B$. The dash-dotted horizontal line indicates the prediction of the CSR for the ratio $a_1/a_2$, whereas the solid grey horizontal line illustrates the theoretical prediction of the CSR for $a_3$ according to Eq.~(\ref{eq:kappa_lambda}). The fits were performed to the subset of simulations shown in Fig.~\ref{fig:sim_vs_theory_loglog} that match the turbulence level $b/B$ to which the points correspond in the present plot.}\label{fig:fit_bB}%
\end{figure}

To further investigate the turbulence-level dependence of the agreement between theory and simulations, we parameterize the CSR as
\begin{align}\label{eq:fit_large}
\frac{\kappa_\perp}{\kappa_\parallel} = \frac{a_1}{1+a_2 (\lambda_\parallel/r_\mathrm{g})^{a_3}} \stackrel{\lambda_\parallel \gg r_\mathrm{g}}{\approx}  \frac{a_1}{a_2}\left(\frac{\lambda_\parallel}{r_\mathrm{g}}\right)^{-a_3},
\end{align}
where the latter expression yields as a good approximation in the limit of large $\lambda_\parallel$, and fit the simulation data for turbulence levels. The best-fit results for the parameters $a_1, a_2$, and $a_3$ for all turbulence levels presented in this study are shown as functions of $b/B$ in Fig.~\ref{fig:fit_bB}. The CSR predictions are $a_1 =1$, $a_2 = 1$, and $a_3 = 2$ as indicated by the horizontal solid lines in the plot. As it is apparent in Fig.~\ref{fig:sim_vs_theory_loglog}, the approximation presented in Eq.~(\ref{eq:fit_large}) for large $\lambda_\parallel$ is used for the fits for small turbulence levels, where all simulation points approximately obey $\kappa_\perp/\kappa_\parallel \propto (\lambda_\parallel / r_\mathrm{g})^{a_3}$. Theoretically, this criterion is justified by the fact that the parallel diffusion coefficients scale with $(b/B)^\alpha$ (with $\alpha = -2$ for QLT). This scaling is directly transferable to the parallel mean-free paths, since $\lambda_\parallel = 3\kappa_\parallel/c$. Lower turbulence levels thus lead to larger $\lambda_\parallel$. Tests have shown the value of 2 for the turbulence level to be practical as a switch between both expressions presented in Eq.~(\ref{eq:fit_large}).

For high turbulence levels, the fits agree well with these CSR predictions. A difficulty arises at small turbulence levels, where, due to the dependence of the parallel diffusion on the turbulence level, simulation data only exist for $\lambda_\parallel \gg r_\mathrm{g}$. This leads to large uncertainties in the fits. As we commented below Eq.~(\ref{eq:error}), in these cases there may also be a contribution from FLRW.

\subsection{Implications for time scales of diffusion}
In Section \ref{sec:3}, we found that the classical scattering relation (CSR) describes the diffusion coefficients accurately for a wide range of parameters. This is beneficial in that one may recover the difficult-to-evaluate perpendicular diffusion from the better-understood and more accessible parallel diffusion using only a few, physically well-motivated assumptions. This finding 
is complemented by our results on the isotropy of the diffusion time scales. We have shown that the diffusion time scales coincide in both parallel and perpendicular directions over a very large parameter space, even if the diffusion coefficients are anisotropic ($\kappa_\perp \ll \kappa_\parallel$) due to strong coherent background fields. The isotropy in the diffusion time scales is caused by two effects:
\begin{enumerate}
    \item Increasing the turbulence level isotropizes the diffusion process and therefore also the time scales needed to correctly describe transport diffusively in the parallel and perpendicular directions. This manifests itself in the fact that the CSR is able to describe the data more accurately with increasing turbulence level, independent of energy, as this theory applies $\lambda_\perp = \lambda_\parallel$.
    \item Isotropization of the diffusion time scales also arises for increased particle energy, especially for $r_\mathrm{g} \gtrsim l_\mathrm{c}$, as the contribution from resonant scattering decreases. It is important to emphasize that equal parallel and perpendicular time scales for diffusion do not lead to equal diffusion coefficient components for this case. By increasing the energy, the parallel mean-free path increases as $\lambda_\parallel \propto E^2$ and suppresses perpendicular diffusion according to 
    \begin{align}
        \kappa_\perp = \frac{\kappa_\parallel}{1+\lambda_\parallel^2 /r_\mathrm{g}^2} \stackrel{\lambda_\parallel \gg r_\mathrm{g}}{\approx} \frac{c\, r_g^2}{3 \lambda_\parallel}.
    \end{align}
    While $\lambda_\perp = \lambda_\parallel \propto \kappa_\parallel  \propto E^2$, the perpendicular diffusion coefficient is energy-independent.
    Note that a related setup involves the spatial diffusion of beam-injected ions and alpha particles in fusion plasmas, where energy-independent diffusion due to magnetic fluctuations is predicted by decorrelation theory \citep{Hauff09,Pueschel12,Pueschel12a}. Future work, accounting for pitch-angle-dependence, will assess to which degree this theory can be seen as a (non-relativistic) extension of the present results.
    
    The finding of isotropic diffusion time scales in combination with energy-independent perpendicular diffusion indicates that high-energy particles follow field lines, a behaviour commonly seen for low-energy particles with small gyroradii. The identification of field-line random walk as the cause for diffusion of high-energy particles is in agreement with \citet{Minnie2009}.
\end{enumerate}

Here, it should be noted that the presented findings are only valid for acceleration times smaller than the escape times of the particles; otherwise, particles would already become diffusive during acceleration. When diffusion times are smaller than acceleration times, these results have an important consequence for the escape times perpendicular to the ordered field. Assuming transport to be in the diffusive limit, the escape time of charged particles to reach distance $d$ yields
\begin{align}\label{eq:esc}
    \tau_{\mathrm{esc},\perp} \approx \frac{d^2}{2\kappa_\perp} \stackrel{\lambda_\parallel \gg r_\mathrm{g}}{\approx} \frac{3d^2\lambda_\parallel}{2cr_\mathrm{g}^2},
\end{align}
which again is energy-independent under the scaling $\lambda_\parallel \propto r_\mathrm{g}^2$ for high-energy particles $l_\mathrm{c} \lesssim r_\mathrm{g}$. The time scale needed to reach the diffusive limit is given by \citep{2021arXiv210711386R}
\begin{align}
    \tau_{\mathrm{diff},\perp} \approx \frac{\lambda_\perp }{c} \approx \frac{\lambda_\parallel}{c}.
\end{align}
For $l_\mathrm{c} \lesssim r_\mathrm{g}$, the diffusive assumption used to derive the perpendicular escape time can be seen to break down when computing the ratio of the time scales for diffusion and escape
\begin{align}
    \frac{\tau_{\mathrm{diff},\perp}}{\tau_{\mathrm{esc},\perp}} \approx \frac{2r_\mathrm{g}^2}{3d^2}.
\end{align}
For the diffusive approach to be applicable, this ratio must be less than 1, from which the condition $r_\mathrm{g} \lesssim \sqrt{3/2} d$ follows. This condition for $r_\mathrm{g}$ should only be understood as an upper limit since in perpendicular diffusion, there is the peculiarity that the maximum of the perpendicular running diffusion coefficient in weak turbulence is already reached in the first gyration (max($\kappa_{\perp}(t)) = c\,r_\mathrm{g} / \pi)$ and does not coincide with the final converged $\kappa_\perp$. Using the temporal maximum of the perpendicular diffusion coefficient as a more realistic approach in Eq.~(\ref{eq:esc}) for the goal of deriving a realistic condition for the validity of the perpendicular escape time results in 
\begin{align}\label{eq:esc2}
    \tau_{\mathrm{esc},\perp} \approx \frac{d^2}{2\kappa_\perp} \approx \frac{d^2\,\pi}{2cr_\mathrm{g}},
\end{align}
and therefore
\begin{align}
    \frac{\tau_{\mathrm{diff},\perp}}{\tau_{\mathrm{esc},\perp}} \approx \frac{2r_\mathrm{g} \lambda_\parallel}{3\pi\,d^2}.
\end{align}
Interpreting requirements more strictly, the diffusive limit should only be applied under the condition 
\begin{align}\label{eq:criterion}
r_\mathrm{g} \lesssim \frac{3\pi d^2}{2\lambda_\parallel}.
\end{align}
The practicality of this criterion is attributed to the fact that it depends only on well-studied parallel diffusion, making it easy to evaluate. Especially as high-energy particles in weak turbulence have large parallel mean-free paths $\lambda_\parallel$, Eq.~(\ref{eq:criterion}) poses a restrictive criterion for the usage of the perpendicular escape times based on the perpendicular diffusion coefficient.

This criterion can be used to check if cosmic rays have enough time to become diffusive in astrophysical environments. The condition becomes particularly useful and at the same time restrictive for high energy particles (in the QBR) which are in compact structures with small values for $d$. In this context, examples include perpendicular diffusion in elongated jets with relatively small cross-sectional scales with respect to the extent along the jet axis.

\section{Conclusion}
Diffusive propagation and its degree of anisotropy is essential for modelling cosmic-ray transport in turbulence. In this study, we examine the ratios of perpendicular to parallel diffusion coefficients in isotropic three-dimensional Kolmogorov turbulence superimposed on a uniform background field over a parameter space spanning from the resonant scattering transport regime to the quasi-ballistic regime, for weak (QLT limit) and strong turbulence levels ($b/B \gtrsim 1$).
Test-particle simulations in synthetic magnetic fields and analysis as defined by the Taylor-Green-Kubo formalism are the basis of this study.
We show that a simple analytical approach, the Classical Scattering Relation (CSR), yields a very good description of the simulation results throughout a large portion of parameter space, which covers several transport regimes and turbulence levels. This includes high-energy particles in the quasi-ballistic regime with large mean-free paths that follow field lines and exhibit energy-independent perpendicular diffusion coefficients that are characteristic of FLRW diffusion. Where resonant scattering is dominant over diffusive transport and the influence of Field Line Random Walk (FLRW) is subordinate, however, significant discrepancies between CSR predictions and our simulation results arise. These discrepancies are only apparent in the resonant scattering regime for weak turbulence levels. For all other parameter regimes, good agreement between CSR and simulations is found, from which the following findings emerge
\begin{enumerate}
    \item the perpendicular diffusion coefficients follow directly from the parallel component via \begin{align}\label{eq:kappa_perp_final}
        \kappa_\perp = \frac{\kappa_\parallel}{1+(3\kappa_\parallel)^2/(c\,r_\mathrm{g})^2}, 
    \end{align}
    and
    \item the diffusion time scales for parallel and perpendicular diffusion are identical and can be obtained directly from the parallel diffusion coefficients via $\tau_{\mathrm{diff},\perp} \approx \tau_{\mathrm{diff},\perp} \approx 3\kappa_\parallel/c^2$.
\end{enumerate}
Since the denominator in Eq.~(\ref{eq:kappa_perp_final}) is greater than or equal to one for all parameter combinations (and much greater for weak turbulence levels), the parallel diffusion coefficients are greater than the perpendicular counterparts. Viewed in the context of finding (ii), Eq.~(\ref{eq:kappa_perp_final}) provides crucial insights into when particles may be considered diffusive. Whereas the time scales of parallel diffusion are proportional to the parallel diffusion coefficient, the time scales of perpendicular diffusion increase as the perpendicular diffusion coefficient decreases. In exploring this behaviour, we preset a condition, see Eq.~(\ref{eq:criterion}), on the time scale of perpendicular diffusion permitting estimation of whether diffusion perpendicular to the directed magnetic field can occur in given astrophysical environments. We anticipate that this criterion will be especially useful in describing the propagation of high energy cosmic rays which interact primarily with large scale isotropic turbulence in the Galactic disk and beyond.

\section*{Declarations}

\subsection*{Funding}
This work is supported by the ``ADI 2019’’ project funded by the IDEX Paris-Saclay, ANR-11-IDEX-0003-02 (PR). PR also acknowledges support by the German Academic Exchange Service and by the RUB Research School via the \textit{Project International} funding. LM acknowledges financial support from the Austrian Science Fund (FWF) under grant agreement number I 4144-N27. EZ gratefully acknowledges support by the US National Science Foundation through grant AST2007323.
 
\subsection*{Acknowledgments}
\noindent We thank Leander Schlegel for usefull discussions.

\subsection*{Availability of data}
The simulation data presented in this paper are available to interested researchers upon reasonable request.

\subsection*{Code availability}
Simulations were performed with the publicly available tool CRPropa \citep{CRPropa2017} (the specific version used for the simulations is CRPropa 3.1-f6f818d36a64), supported by various analysis tools: NumPy~\citep{van_der_Walt_2011}, Matplotlib~\citep{Hunter_2007}, Pandas~\citep{pandas}, and jupyter notebooks~\citep{jupyter-notebook}.





\bsp	
\label{lastpage}

\end{document}